# Are We Standing on Unreliable Shoulders? The Effect of Retracted Papers Citations on Previous and Subsequent Published Papers: A Study of the Web of Science Database

**Sepideh Fahimifar**
Assistant prof., Department of Information and Knowledge Science, Faculty of Management, University of Tehran, Tehran, Iran
Corresponding Author: sfahimifar@ut.ac.ir
ORCID iD: https://orcid.org/0000-0001-5182-9159

**Ali Ghorbi**
Master of Scientometrics, Faculty of Management, University of Tehran, Tehran, Iran.
alighorbi73@ut.ac.ir
ORCID iD: https://orcid.org/0000-0003-1411-575X

**Marcel Ausloos**
School of Business, University of Leicester, Brookfield, Leicester, United Kingdom
GRAPES, Liege Angleur, Belgium
Department of Statistics and Econometrics, Bucharest University of Economic Studies, Calea Dorobantilor, Bucharest, Romania
marcel.ausloos@uliege.be
ORCID iD; https://orcid.org/0000-0001-9973-0019

**Abstract**

The present research attempts to identify the impact of retracted papers on previous or subsequent papers. We consider the 5693 retracted papers from 1975 to 2020 indexed in the Web of Science database based on bibliometric methods. We use HistCite, Excel, and SPSS software as technical means. The findings suggest a significant difference between the average number of retracted and unretracted papers when cited in retracted papers. Furthermore, there is a significant difference between the average number of unretracted and retracted papers citing retracted papers. The reasons for the retraction of an article may not be the previous retracted papers, yet unretracted papers may be retracted later because of referring to (many) retracted papers. It is deduced that proprietors of citation databases should carefully focus on these papers by checking references to each new paper citing previously retracted papers.

## Introduction

Scholarly journal publications continue to mark the state of progress within a research community (Shuai, Rollins, Moulinier, Custis, Edmunds & Schilder, 2017). However, publishing an academic paper sometimes deviates from its original path. It ends in retraction, the most serious stigma in scientific research (Chen, Hu, Milbank & Schultz, 2013) beside coercive citations leading to anomalous measures of research quality (Herteliu, Ausloos,

Ileanu, Rotundo & Andrei, 2017) and plagiarism (Sharma & Singh, 2011). Retraction is a public notice that an article should be withdrawn because of errors or unsubstantiated data. Rather alarmingly, the number of retracted papers has significantly increased (He, 2013) recently. Studying papers about the retraction subject, we can state that research on retracted papers has mainly examined:

1. The retraction reasons: due to publisher's error, authors' error, previously un-accounted-for error, self-citations, honorary writing (bringing in the names of those who have not played a role or at least played a minor role) (Budd, Sievert & Schultz, 1998; Foo, 2011; Steen, 2011a; Grieneisen & Zhang, 2012; Marcus & Oransky, 2014; Moylan & Kowalczuk, 2016; Dal-Ré & Ayuso, 2019).

2. The enhanced rate of retracted papers (Tchao, 2014; Brainard, 2018; McCook, 2018; Al-Ghareebet al., 2018) that the reason for this kind of behavior seems unclear (He, 2013; Fanelli, 2009; Steen, Casadevall, & Fang, 2013); some reasons are related to "economic aspects", inherent to the present system of science development such as grants, prizes, funding, jobs, etc. (Fang & Casadevall, 2011; Stern, Casadevall, Steen & Fang, 2014). On the other hand, the rapid rate of discovering retracted papers has a root in the self-expression of authors (Wager & Williams, 2011), the creation of some websites in order to track research, such as retraction watch[1] (Marcus & Oransky, 2014), an increase in the level of scrutiny over articles by the widespread use of the Internet, and also an in-depth review by some editors-in-chief of the journals or by reviewers (Moylan & Kowalczuk, 2016).

3. The impact of retracted papers on future scientific findings, through examining different kinds of citations in which most citations have been reported as being positive (without any criticism) (Decullier, Huot, Samson & Maisonneuve, 2013; Furman, Jensen & Murray, 2012; Nath, Marcus & Druss, 2006; Trikalinos, Evangelou & Ioannidis 2008).

4. The time lag between the publication of an article and its retraction has also been considered (Chen et al., 2013; Foo, 2011; Steen, Casadevall & Fang, 2013). It takes about two years, "on average", for a paper to be retracted after publication (Bar-Ilan & Halevi, 2017; Budd, Sievert, Schultz & Scoville, 1999; Korpela, 2010; Steen, 2011b; Trikalinos et al., 2008).

5. Effects of citation to retracted papers have been one of the most debated subjects. The evidence suggests that citations of retracted papers continue as if they are valid works (Pfeifer & Snodgrass, 1990; Fang & Casadevall, 2011). Interestingly, citing these papers can have a positive, negative, or neutral effect (Bar-Ilan & Halevi, 2017) before or after a paper's retraction (Campanario, 2000; Couzin & Unger 2006). However, editorial and publisher policies may differ and change with time (Teixeira da Silva & Dobránszki, 2017a). Positive effects are considered when citations use the content of retracted papers to support an argument regardless of the scientific validity of these retracted papers. Negative effects, on the contrary, occur when citations criticize the content of the retracted papers (pre-or post-retraction) and highlight the errors that have occurred in those papers. Of course, neutral effects are for citations that refer to a retracted (or to be retracted) paper in the literature section without any judgment on its validity. Bar-Ilan and Halevi (2017) showed that 83% of citations to retracted papers were positive, 12% were neutral, and only 5% were negative.

Bar-Ilan and Halevi (2017) also suggested that "positive citations" of retracted papers continue to be made despite retraction notice on publisher's online platforms and without attention for reasons of retraction. Moreover, given that it usually takes two years for a paper to be retracted after its publication (Bar-Ilan & Halevi, 2017; Budd et al., 1999; Korpela, 2010; Steen, 2011b; Trikalinos et al., 2008), there is a good chance for a retracted paper to affect its subsequent papers. Additionally, many retracted papers likely continue to receive post-retraction citations (Chen et al., 2013). For example, many retracted articles in radiation oncology appear to be cited after their retraction; plenty has been referenced as legitimate articles (Hamilton, 2019). Another study on retracted and unretracted articles in the engineering domain shows that their citations continue and affect their scientific creditability (Rubbo, Pilatti & Picinin, 2019). Besides, some risks occur after continuing to cite the retracted papers. Some authors believe that their findings and results are still valid (da Silva & Dobranszki, 2017), and this, in some sense, can misguide readers.

---

[1] http://retractiondatabase.org/RetractionSearch.aspx?

It means that these kinds of articles written with falsification, fabrication, or unreliable data may lead to dubious results. For these results, advancement is somewhat deceptive when we use them as a reference or in solving problems. Thus, quantitative knowledge of the number of retracted papers cited in subsequent publications can provide an alarm point to our scientific community and the general public. Besides the probability of retraction study, another purpose of our article is to find (and display) the cause(s) and effect(s) of a retracted paper on subsequent publications. We consider that in publishing such considerations, we tend to be avoiding the insertion of bibliographic information of retracted papers in unretracted articles; in doing so, we hope that we can help increase the growth of deprived science. Our study, therefore, aims to measure the effect of retracted papers on research publications. Our investigation bears upon cases in journals indexed in the WoS database. It is the most important database globally, covering various disciplines and enabling comparisons across scientific areas.

First, we observe the timeline of the number of retracted papers. Given the plethora of newly appearing journals, following the open access scheme, we add among "Research Accountability" considerations a study measuring whether there might be a significant correlation between a "journal quartiles" and the number of retracted papers. Measuring the correlation between the "quality" of journals and the number of retracted papers stems from the fact that consequences on research activity might be more drastic if retracted papers are from top journals since most researchers and audiences may rely on their conclusions. Indeed, there is no need to stress that any current research is very dependent on prior research. As a result, earlier papers play a remarkable role in forming those published later. Thus, it is interesting to examine how often the retracted papers are cited and which ones are cited by (to be) retracted papers, whence they influence future research!

Therefore, the present study aims to answer the following questions:
- How many papers have been retracted (from 1975 to 2020)?
- What is the distribution of retracted papers in the journals indexed in the Journal Citation Report database (JCR)?
- What is the effect of retracted papers on JCR journals' papers (retracted or not)?

We also want to achieve some research objectives, including:

1. Investigate whether a significant correlation exists between journal quartiles (Q1, Q2, Q3, and Q4) and the number of retracted papers, i.e., whether retracted papers have significantly been published in high-quality journals with high impact factors.

2. Investigating whether a significant difference exists between the average number of unretracted and of retracted papers citing retracted papers; as a matter of fact, we want to know whether or not the retracted papers are more often cited than retracted paper**s** in comparison with unretracted papers.

3. Investigate whether a significant difference exists between the average number of retracted and unretracted papers cited by retracted papers; i.e., the retracted papers result from previous unretracted papers or other retracted papers.

## Methodology

We collected data from the WoS database (on 7 June 2020)[2] according to the search strategy outlined in Figure 1. The data included total records and cited references between 1975 and 2020. Retraction notices were excluded in order to avoid duplication.

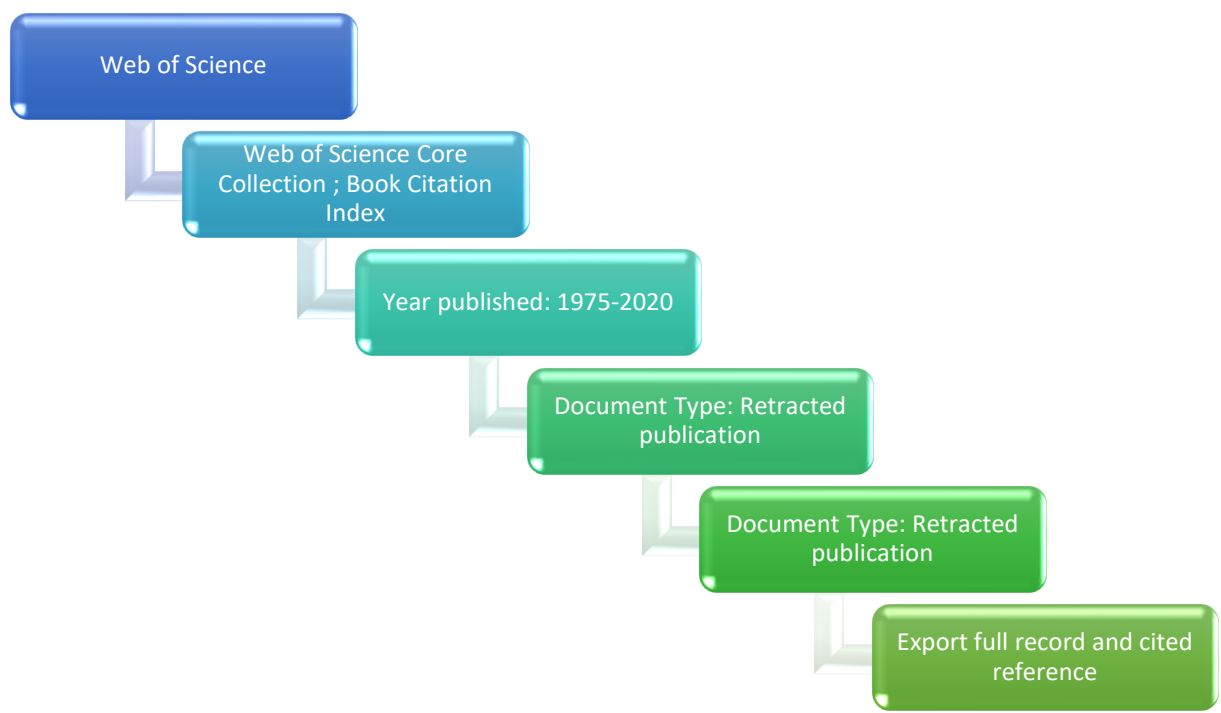

*Figure 1:* Search strategy for collecting data

We should notice that WoS does not have all retracted articles indexed under the publication type "Retracted Publication". Based on WoS, a retracted publication is an article that an author, institution, editor has withdrawn, or a publisher because of errors or unsubstantiated data. However, we use the data from WoS instead of the Retraction Watch Database (retractiondatabase.org) because of our software. In so doing, when gathering data from citation indexes such as WoS and Scopus, we can import data in scientometrics and bibliometrics software. We do not use Retraction Watch because most bibliometrics software only accepts output from some compliant citation databases. We also ignore records with the tag "Retraction" in this study. These include some records which were tagged as retractions in WoS, but the article was not retracted, i.e., practically removed from the journal or website. An example is one article entitled 'Evidence-based indicator approach to guide preliminary environmental impact assessments of hydropower development'. Some editors prematurely published the mentioned article, but the correct version of this article was published in another journal. Moreover, there are some records with 'retraction' tags without any traces of the full text of the paper even with 'retracted paper' captions. Notice that some retraction notices can be found in WoS; however, the full text of the related articles is unexpectedly not tagged as "retracted paper''. An example is ''MicroRNA-493 Suppresses Tumor Growth, Invasion and Metastasis of Lung cancer by Regulating E2F1''.[3]

---

[2] The WOS database was queried for allowing us to retrieve all publications published between 1975 and 2020 without any limitation based on the document type.

We found 5693 retracted papers by 19482 authors published in 2046 journals. After exporting data from WoS as a txt file, we provided data with the editing of the first line of txt file and replaced it with the phrase "ISI Export Format". Then, we imported collected data from WoS into HistCite software. Then, we exported Outputs from the *HistCite*™ software into Excel format as a ".csv" file. However, by using the filtering header and observing the content of each cell, we understood that some data were not correctly sorted out on transfer. As a result, we checked records one by one to ensure that the data was listed correctly, according to

(i) The number of authors of each retracted paper;
(ii) The number of citations to each paper in the entire WoS database (GCS)[4];
(iii) The number of citations to each retracted paper by other retracted papers (LCS)[5];
(iv) The number of unretracted and retracted papers in the reference section of each retracted paper.

The extracted data were then entered into *SPSS* software (version 21.0; SPSS) for statistical analysis.

Next, to answer the research question as to the "quality" of journals in which there were retracted papers, a list of journals along with quartiles published by the JCR database was extracted. Then, we deleted the duplicated titles. Indeed, a journal can be categorized under different quartiles due to the different research areas it covers. We exported the title of journals related to retracted papers as a CSV file. Finally, we compared titles of journals and lists of journals with quartiles to assign a unique quartile to each journal such that the journal with the highest quartile is retained.

## Results

### The number of retracted papers per year

Figure 2 illustrates how many papers were retracted from 1975 to 2020. The number of retracted papers has increased over the period. Firstly, it remained relatively stable from 1975 to 1990. However, after a fluctuation between 1991 and 1996, one can see a sharp increase from 1997 to 2007. While the number of retracted papers fluctuated again between 2008 and 2013, it peaked at 453 retracted papers. After that, the number has decreased but has not much reduced in recent years: it was still above 50 in 2019. In the 21st century, the number of retracted papers has been increasing, the rate roughly doubled from 2003 to 2010, but since

---

[3] In order not to increase the visibility of such articles, the reader can observe that we do not explicitly refer to these in the reference list.
[4] Global Citation Score
[5] Local Citation Score

2015, the number has leveled off (Figure 2).

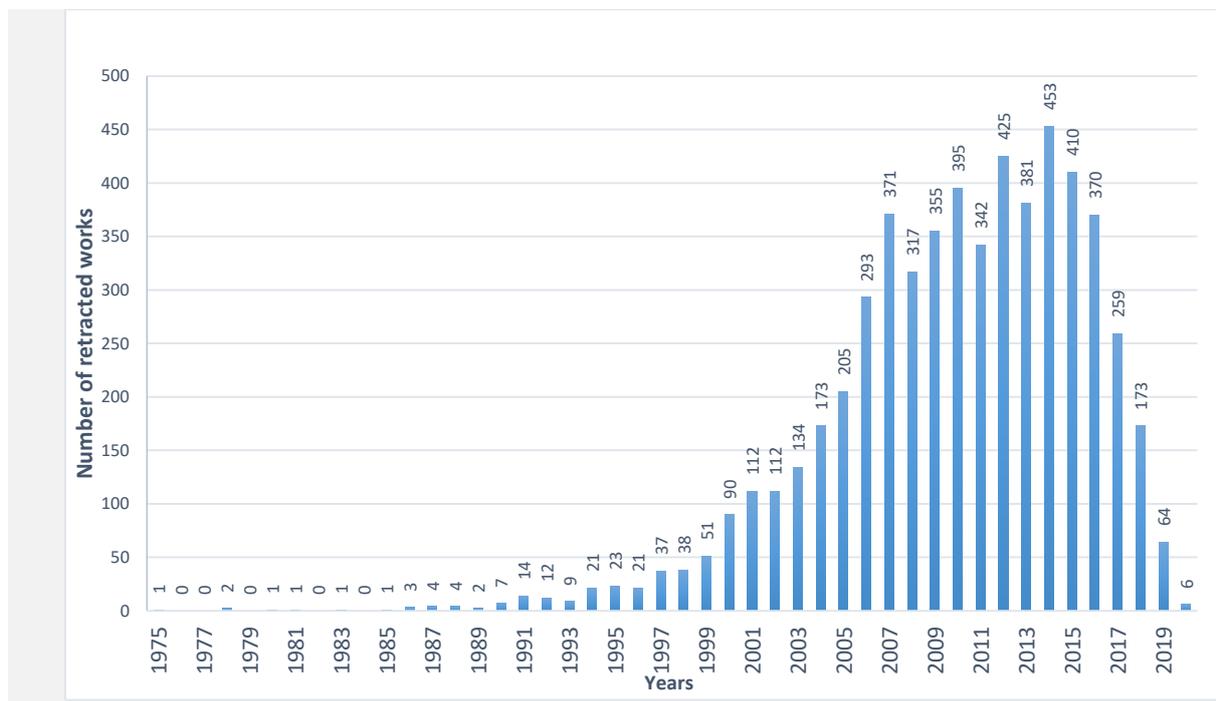

*Figure 2:* The number of retracted papers per year in journals indexed in the WoS

**Distribution of retracted papers in the journals indexed in the JCR database**

Among the 2046 journals[6] with at least one retracted paper, 1893 journals have an impact factor and are indexed in Journal Citation Report; only 153 journals with 495 retracted papers have no impact factor. In the top two quartiles, there are 1278 journals which include 3975 retracted papers. The results of Table 1 suggest that most retracted papers were published in high-impact factor journals.

Table 1

*Dissemination of retracted papers based on Journal ranking in JCR quartiles*

| Total number of retracted papers: 5693 | | | | | | | | |
|---|---|---|---|---|---|---|---|---|
| Total number of retracted papers in journals with impact factor: 5198 | | | | | | | | |
| Journals' quartiles | Q[1] | | Q2 | | Q3 | | Q4 | |
| Frequency and parentage | n | % | n | % | n | % | n | % |
| Retracted papers in journals with impact factor | 2406 | 46% | 1569 | 30% | 753 | 15% | 470 | 9% |
| Total number of journals with retracted papers: 2046 | | | | | | | | |
| Total number of journals with impact factor having retracted papers: 1893 | | | | | | | | |
| Journals' quartiles | Q1 | | Q2 | | Q3 | | Q4 | |

---

[6] Remember that we search whether a journal appears in different quartiles due to its scientific area coverage, and we select the best quartile for such a journal.

| Frequency and parentage | n | % | n | % | n | % | n | % |
|---|---|---|---|---|---|---|---|---|
| Journals with impact factor having retracted papers | 736 | 39% | 542 | 29% | 386 | 20% | 229 | 12% |
| Journals' quartiles | Q1 | | Q2 | | Q3 | | Q4 | |
| Percentage of retracted papers per journal with impact factor (measured for the number of retracted papers from a journal in a given quartile) | 31% | | 35% | | 51% | | 48% | |

This leads to the next section in which we examine whether there is a correlation between the quality of the journals, measured by their quartile indicator, and the number of retracted papers in them.

**Correlation between journal quartiles and the number of retracted papers**

In order to observe whether there is a significant correlation between the journal quartile and the number of retracted papers, we first calculate the number of retracted papers in each journal. As mentioned above, if the journal appeared in more than one discipline, only the highest journal quartile was used. The Spearman correlation test is used for the ordinal and scale variables. The positive correlation coefficient (See Table 2) denotes a significant positive linear correlation between the rank of each journal and the number of retracted papers, even though this correlation seems very weak[7]. Thus, with a 99% confidence level, it can be concluded that the first hypothesis is accepted ($r_s$ [1893] = 0.136, p<0.01).

Table 2

*Correlation between journal quartiles and retracted papers*

| Variable | Number | Median | Standard Deviation | Number of articles in each journal | Quartile indicator of each journal |
|---|---|---|---|---|---|
| 1. Number of articles in each journal | 1,893 | 2.75 | 5.67 | — | 0.136[*] |
| 2. Quartile indicator of each journal | 1,893 | 2.94 | 1.036 | 0.136[*] | — |

*p* < .01.

Correlation between the average number of unretracted and retracted papers citing retracted papers

Let us examine whether retracted papers mainly refer to other prior retracted papers or unretracted papers. In other words, let us examine whether there is any significant difference between the average number of unretracted and that of retracted papers citing retracted papers. In order to do so, we use the Mann-Whitney U test as usually done when one wants to compare the means between two independent groups when the dependent variable is either ordinal or continuous but not normally distributed (Graczyk & Duarte QueiroÂs, 2016). In order to do so, the number of citations of each retracted paper in the WoS database is calculated. This results in the total citation per paper (GCS) and the total number of citations to a retracted paper by other retracted papers (LCS). We subtract the number of retracted papers citations (LCS) from the total number of citations (GCS) in view of calculating the number of citations to a retracted paper by unretracted papers. First, we test the normalization of the data. The result of the Kolmogorov-Smirnov test shows that the data (citing retracted and unretracted papers) is not normally distributed because the P-value is less than 0.01 (See Table 3). Therefore, one should use a Mann-Whitney U test instead of an Independent t-test.

---

[7] Correlation is the measure of how two or more variables are related to one another. It is a range from 0 to 1 including 00-.19 "very weak", 20-.39 "weak", 40-.59 "moderate", .60-.79 "strong", and .80-1.0 "very strong".

Table 3
*Tests of normality distribution for retracted and unretracted papers*

| Variable | *df* | *P-value* |
|---|---|---|
| 1. Cited retracted papers | 5693 | 0 |
| 2. Cited unretracted papers | 5693 | 0 |

The Mann-Whitney U test (See Table 4) indicates that the number of citations received by the retracted papers from unretracted papers is greater than that from the other retracted papers (see mean rank). Therefore, there is strong evidence to support a difference between citations received by retracted papers from both non-retracted and other retracted papers. Given the fact that the number of papers citing retracted papers is greater than the number of papers citing unretracted papers, it can be argued that there is a "chance" for these citing papers to be considered retracted in the future. Therefore, if the theoretical and substantive discussions within unretracted papers originate from the content of retracted papers, those discussions can be considered "potentially unreliable". Nevertheless, for a satisfactory conclusion, one should analyze the text content to determine whether the citation is a criticism or a positive appreciation.

Table 4
*The significant mean difference between the citing retracted and unretracted papers*

| Variable | *Number* | *Mean rank* | *P-value* |
|---|---|---|---|
| 1. Cited retracted papers | 5693 | 3046.86 | 0 |
| 2. Cited unretracted papers | 5693 | 8340.14 | 0 |

**Difference between the average number of retracted and unretracted papers cited by retracted papers**

This section examines whether authors of retracted papers are mostly referring to other previously retracted papers or have mostly mentioned unretracted papers in their publications. Whether there is any significant difference between the average numbers of unretracted and retracted papers cited by retracted papers (i.e., retracted papers cite other papers) is again calculated through a Mann-Whitney U test.

Hence, we subtract the number of retracted references from the total number of references for calculating the number of unretracted references. First of all, because of scale data, we calculated the normalization by using the Kolmogorov-Smirnov test (See Table 5).

Table 5

*Test of normality of the number of references for each retracted paper*

| Variable | df | P-value |
|---|---|---|
| 1. Cited retracted paper | 5693 | 0 |
| 2. Cited unretracted papers | 5693 | 0 |

The Mann-Whitney U test indicates that the greater number of references cited by retracted papers are unretracted papers (See Table 6).

**Table 6**

*The significant mean difference between the number of cited retracted and unretracted papers*

| Variable | n | Mean rank | P-value |
|---|---|---|---|
| 1. Cited retracted paper | 5693 | 2894.95 | 0 |
| 2. Cited unretracted papers | 5693 | 8492.05 | 0 |

**Discussion**

This research investigated the number of retracted papers per year in the WoS database over a 45-year interval, in particular, to examine whether there is a relationship between journal quartile and the number of retracted papers in the WoS listed journals as well as if there is any relationship between retracted papers and the papers they cite and the papers in which they are cited.

According to our findings, the number of retracted papers has increased over the examined period. There is nowadays quite a decrease following the prominent peak in 2015. It seems that the detection of retractable papers has been recently more effective due to the advancement of fraud recognition systems and more scrutiny in the peer-reviewing process, and, *in fine*, pressure from peers. As a result, invalid science is being identified faster than ever before (Bornemann-Cimenti, Szilagyi, & Sandner-Kiesling, 2016). The increasing number of retracted papers should be of concern due to the negative impact such a retraction has; sometimes, there can be dangerous consequences due to the application of retracted papers that had announced findings in the real world, specifically in health-related disciplines[8]. This aspect becomes very critical, especially when retraction is misconduct or fraud. It is unfortunate but true that an enormous number of papers have been retracted in recent years due to research misconduct (Fang, Steen & Casadevall, 2012; Marcus & Oransky, 2014; Samp, Schumock & Pickard, 2012; van Noorden, 2011).

---

[8] The case of the hydroxyhlorine potential effect as a drug against covid-19 is of world knowledge, at the time of writing this paper.

The current study's findings show that the largest number of retracted scientific papers is found in journals with high impact factors (Q1). Our findings are consistent with those of Steen (2011c) based on the data extracted from the PubMed database. Similarly, Fang et al. (2012) also indicated a significant correlation between the impact factor of journals and fraud or honest errors.  Rubbo, Helmann, Bilynkievycz dos Santos & Pilatti (2019) also showed that journals with high impact factors usually have more retracted papers than journals with low impact factors in the engineering field. Moreover, we observed a significant correlation between the journal quartiles and the number of retracted papers. In an investigation into the retraction policy of high-impact-factor journals indexed in the JCR 1999 database, Atlas (2004) reported that 3.27% of journals  (i.e., 4 out of 122) had a retraction policy. Thus, it would be better for journal editors to add the retraction policy on the journal website to inform their authors. Moreover, some examples of retraction can be helpful for authors because these examples could clearly show how their articles may be retracted.

The analysis of the second hypothesis suggests that retracted papers are among the citations of a significant number of unretracted papers. This would result in the likeliness of the scientific basis of published papers being unconsciously and unknowingly derived from the false claims of retracted papers. Identifying this type of effect is more complicated than recognizing articles and scientific papers that have been retracted (Chen et al., 2013). This is especially important in sensitive areas such as medicine, veterinary medicine, and plant medicine associated with life-related issues. While it is acceptable that researchers use and refer to retracted papers as unsuitable examples and dangerous practices of knowledge dissemination, it is also recommended that they do not include these papers in the references section of their papers. Instead, it is suggested that these papers are only referenced in footnotes (Eisenach, 2009; Cosentino & Veríssimo, 2016).

The analysis of the third hypothesis suggests that retracted papers have cited unretracted papers more often than retracted papers. It would seem that retracted papers must not have led to the production of subsequent retracted papers. The "technical accountability point" about this finding is whether references in retracted papers, which have cited unretracted papers, should be omitted in the measure of the number of unretracted papers' citations or not. From our standpoint, citation databases such as WoS and Scopus should address this problem. Indeed, when one paper is retracted, it is not fair to calculate the number of its cited references as a correct citation in the calculated number of citations of the previous papers because this is a retracted paper. Such references should not count in calculating, for example, the Hirsch index (Teixeira da Silva & Dobránszki, 2017a; 2017b) or when measuring the weight of co-authors (Ausloos, 2015a; 2015b).

Therefore, this subject reinforces the concern in the citation debate of papers that have been recognized as reliable sources. As a strategy to address the issue of citing retracted publications, Teixeira da Silva and Bornemann-Cimenti (2017) state that reference management software should implement a retraction check during the online submission of scholarly articles. This detects any retracted papers that authors might have unknowingly cited in their papers. It can be implemented through modern techniques (Mrowinski, Fronczak, Fronczak, Nedic & Ausloos, 2016; Mrowinski, Fronczak, Fronczak, Ausloos, & Nedic, 2017). We suggest that the number of unretracted papers' citations through retracted papers should also be omitted. Moreover, Oransky and Marcus (2010) and Teixeira da Silva and Bornemann-Cimenti (2017) suggest that the status of a retracted paper should be shown along with the title of the work or be highlighted in a colorful bar near the title. It is noted that not all retraction notices are always transparent and explicit. This leads to retracted papers continuously receiving citations from other papers including unretracted papers. It is, therefore, recommended that the Committee on Publication Ethics (COPE) suggests unique instructions on writing the retraction notice. Van der Vet and Nijveen (2016) also suggest that authors whose papers are detected for including retracted citations should be allowed, even post-publication of their papers, to modify their papers by eliminating the retracted resources.

On the other hand, this procedure may reduce the negative impact of science growth. Given that retracted papers significantly cite unretracted scientific papers, it is suggested that the effects of previous papers be identified on subsequent papers. If this effect is based on the research method or other essential parts of the research, especially in the health sciences and medicine, the subsequently published papers should be aware of and purge themselves from retracted paper adverse effects as quickly as possible.

After all, it should be noted that all retracted papers are not of the same type and occur for several reasons discussed above (Budd et al., 1998; Dal-Ré & Ayuso, 2019). Therefore, the same decision should not be made about all retracted papers. For instance, a retracted article due to figure manipulation could still be considered valid since its methodology remains thorough (Teixeira da Silva & Bornemann-Cimenti, 2017). In that spirit, it is recommended that owners of citation databases develop a retraction index that calculates a weight of retraction for retracted papers based on their type of retraction. Such an index advises a threshold above which papers should not be used in any way. Furthermore, given that many authors may not be familiar with the instances of retraction, it is suggested that journals make available examples of retracted papers on their websites. Many journals, in a section they have as guides for authors, refer the authors to the citation style guides they use and trust that authors follow the guides carefully. This is only useful for plagiarism.

## Conclusion

Many unretracted papers have cited retracted papers. These are not good news because of the possible danger a retracted paper can cause. Thus, the authors of the unretracted papers should evaluate their articles and inform the editors if they have had a notable impact on their results. Journals can also choose to review papers with more scrutiny if authors have had previously retracted papers. In order to do so, journals can create a list of retracted papers' authors to which authors of new submissions are compared before the review process. It may help identify a fraudulent paper in a shorter time since, in conformity with the PubMed database, quite a few retracted papers, roughly 53% of fraudulent papers, were written by a first author who had written other retracted papers as well (Steen, 2011c). It would also be interesting to observe the country (van Leeuwen & Luwel, 2014) and/or institutional affiliation of authors retracting papers (Ahmed, 2020).

There is still a need for deep research to understand whether the reason for the retraction of an article is the use of previously retracted papers or commitment of errors such as plagiarism and data manipulation, whatever the misconduct type (DuBois, Anderson, Chibnall, Carroll & Rubbelke, 2013) beside publisher or author error(s). It seems critical to re-assess the validity of the claims made in the unretracted papers referenced to the retracted papers in case they are required to be retracted as soon as possible so that they will not impact any future paper. As a result, it is suggested that databases oblige journals to monitor papers that use retracted papers as arguments for their report.